\documentclass{article}
\usepackage{spconf,amsmath,graphicx}
\usepackage{amssymb, hyperref}

\title{Segmental DTW: A Parallelizable Alternative to Dynamic Time Warping}
\name{TJ Tsai\thanks{\copyright~2021 IEEE. Personal use of this material is permitted. Permission from IEEE must be obtained for all other uses, in any current or future media, including reprinting/republishing this material for advertising or promotional purposes, creating new collective works, for resale or redistribution to servers or lists, or reuse of any copyrighted component of this work in other works.}}
\address{Harvey Mudd College, Claremont, CA USA}
\begin{document}
\ninept
\maketitle
\begin{abstract}
In this work we explore parallelizable alternatives to DTW for globally aligning two feature sequences.  One of the main practical limitations of DTW is its quadratic computation and memory cost.  Previous works have sought to reduce the computational cost in various ways, such as imposing bands in the cost matrix or using a multiresolution approach.  In this work, we utilize the fact that computation is an abundant resource and focus instead on exploring alternatives that approximate the inherently sequential DTW algorithm with one that is parallelizable.  We describe two variations of an algorithm called Segmental DTW, in which the global cost matrix is broken into smaller sub-matrices, subsequence DTW is performed on each sub-matrix, and the results are used to solve a segment-level dynamic programming problem that specifies a globally optimal alignment path.  We evaluate the proposed alignment algorithms on an audio-audio alignment task using the Chopin Mazurka dataset, and we show that they closely match the performance of regular DTW.  We further demonstrate that almost all of the computations in Segmental DTW are parallelizable, and that one of the variants is unilaterally better than the other for both empirical and theoretical reasons.
\end{abstract}
\section{Introduction}
\label{sec:intro}

This paper explores a parallelizable alternative to dynamic time warping for globally aligning two feature sequences.  Dynamic time warping (DTW) is a standard method for determining the optimal alignment between two sequences.  One of its main limitations is its quadratic computational and memory cost, which limits its use to shorter sequences in many situations.  Furthermore, because the DTW algorithm is inherently sequential, it cannot be parallelized for determining the global alignment between two sequences.  Our goal in this paper is to explore alternatives to DTW that (a) estimate the global alignment between two sequences and (b) are amenable to parallelization.

Many previous works have explored variants of DTW.  Most works generally fall into one of three groups.  The first group are works that explore ways to speed up exact subsequence DTW search.  Some of these approaches include using lower bounds \cite{zhang2011inner}\cite{keogh2009supporting}, early abandoning \cite{rakthanmanon2012searching}\cite{li2007ea}, and utilizing multiple cores \cite{srikanthan2011implementing} or specialized hardware \cite{sart2010accelerating}.  The second group are works that extend the behavior of DTW in various ways.  In the music information retrieval literature, some examples include doing the time warping in an online fashion \cite{macrae2010accurate}\cite{dixon2005live}, handling repeats and jumps \cite{FremereyMC10_RepeatsJumps_ISMIR}\cite{shan2020improved}, handling subsequences or partial alignments \cite{MuellerA08_PathConstrained_ICASSP}\cite{muller2008joint}, handling pitch drift in a capella performances \cite{waloschek2018driftin}, and taking advantage of multiple recordings \cite{wang2016robust}.  The third group are works that mitigate the computation and/or memory cost of DTW by proposing approximations, modifications, or alternative alignment algorithms.  Some examples include imposing fixed constraints like the Sakoe-Chiba band \cite{sakoe1978dynamic}, using a multi-resolution alignment approach \cite{MuellerMK06_EfficientMultiscaleApproach_ISMIR}\cite{SalvadorC04_fastDTW}, and calculating or estimating the alignment path with limited memory \cite{tralie2020exact}\cite{PraetzlichDM16_MsDTW_ICASSP}.\footnote{The very recent work by Tralie and Dempsey \cite{tralie2020exact} computes an exact DTW alignment with linear memory.  They point out that many operations in the algorithm are parallelizable, though their primary focus is on reducing memory and no specific runtimes are reported in the paper.  Furthermore, the parallelizable operations require frequent communication between the threads, so it is unclear how much the theoretical parallelizability translates into actual reduced runtime.  In contrast, our approach parallelizes computation in a way that requires very little communication between threads.}  The approach proposed in this work also falls within this third group, where the focus is on reducing time (expensive) rather than computation (cheap).

This paper explores two variants of a previously proposed algorithm called Segmental DTW \cite{tsai2017make}.  Segmental DTW was originally proposed to solve an entirely different problem: aligning an ordered set of audio segments against a long reference recording.  The focus of the original paper is on handling discontinuities coming from the unknown gaps between the segments.  In this paper, we explore two different variants of Segmental DTW to solve a problem that is arguably much more general: approximating DTW with a parallelizable alternative.  The high-level approach is to break the global cost matrix into several sub-matrices, process the sub-matrices separately, and then combine the results in a way that leads to a globally optimal alignment path.  This approach leads to an alignment algorithm that can approximate DTW and is parallelizable.

This paper has two main contributions.  First, we present two global alignment algorithms that are parallelizable.  One of the algorithms (weakly-ordered Segmental DTW) offers only loose guarantees on the monotonicity of the alignment paths.  The other algorithm (strictly-ordered Segmental DTW) offers strict guarantees on monotonicity, in exchange for additional computation.  Both algorithms can be considered as alternatives to DTW that estimate the global alignment between two sequences of features.  Second, we characterize the behavior of the proposed algorithms on an audio-audio alignment task.  We find that the alignment accuracy closely matches that of regular DTW, and we show that nearly all of the computations can be parallelized.  We further show through empirical and theoretical arguments that, contrary to intuition, the weakly-ordered variant is a unilaterally better alternative to DTW than the strictly-ordered variant.\footnote{Code can be found at \url{https://github.com/tjtsai/SegmentalDTW}.}

\section{System Description}
\label{sec:system}

In this section we describe two variants of Segmental DTW that are parallelizable alternatives to DTW.  For the sake of concreteness, we describe systems that approximate DTW with transitions $\{(1,1), (1,2), (2,1)\}$ and corresponding multiplicative weights $\{2, 3, 3\}$.  This scheme assumes a maximum time warping factor of two, and it weights transitions based on their Manhattan distance.  %Note that this weighting means that all valid alignment paths are equally weighted, since all paths across the cost matrix have the same Manhattan distance.

\subsection{Weakly-ordered Segmental DTW}
\label{subsec:wsdtw}

\begin{figure}
	\centering
	\includegraphics[width=.8\linewidth]{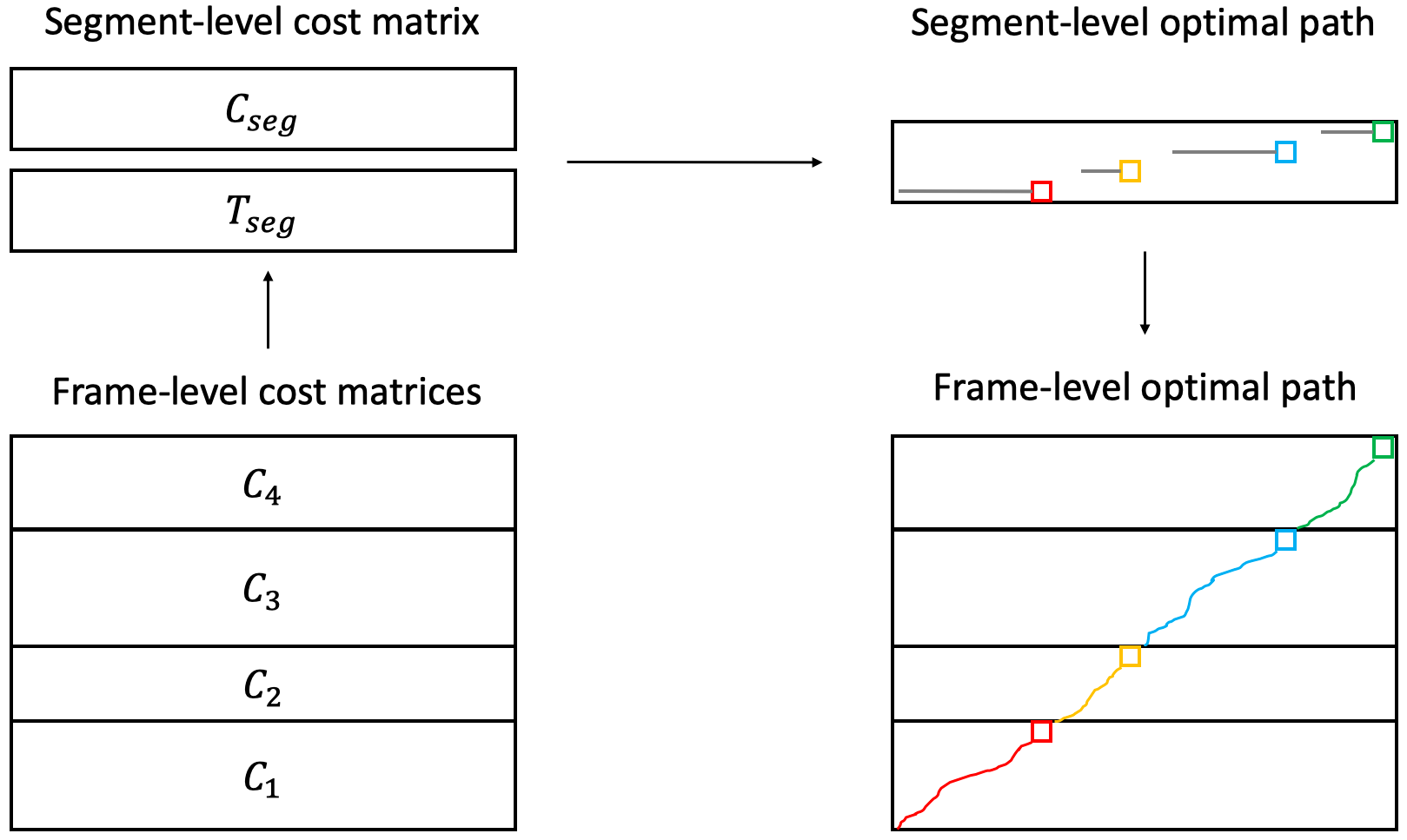}
	\caption{Overview of the four main steps in Segmental DTW.}
	\label{fig:overview}
\end{figure}

The first variant is called weakly-ordered Segmental DTW, which we abbreviate as WSDTW.  This algorithm consists of four main steps, as shown in Figure \ref{fig:overview}.

The first step is to break the global cost matrix $C$ into chunks $C_i$ and perform subsequence DTW on each chunk.  Let the two feature sequences be denoted as $(x_1, x_2,\dots, x_N)$ and $(y_1, y_2, \dots, y_M)$, so that the cost matrix $C$ is an $N \times M$ matrix.  We break up the sequence $(x_1,\dots, x_N)$ into $K$ (approximately) equal size subsequences.  We then perform subsequence DTW using each subsequence as a query and the entire sequence $(y_1, y_2,\dots,y_M)$ as the reference.  Subsequence DTW is a variant of DTW in which, given a short query sequence, the best matching subsequence in a longer reference sequence is found.  Unlike regular DTW, subsequence DTW allows alignment paths to start and end anywhere in the reference sequence without penalty.  We perform the subsequence DTW with transitions $\{(1,1), (1,2), (2,1)\}$ and weights $\{1, 1, 2\}$, where the $(2,1)$ transition corresponds to two steps along the query sequence and one step along the reference sequence.\footnote{Note that, if the three transitions were weighted equally, the algorithm would be incentivized to only take $(2,1)$ steps, since it would accumulate half as many cost elements as a path taking only $(1,1)$ steps.  The reader is referred to \cite{muller2015fundamentals} for more details on subsequence DTW.}  At the end of this first step, we have performed subsequence DTW on each sub-matrix $C_i \in \mathbb{R}^{N/K \times M}$, $i=1,2,\dots,K$ to produce a cumulative cost matrix $D_i$ and backtrace matrix $B_i$, where $D_i$ contains the optimal cumulative path scores and where $B_i$ specifies the optimal steps taken at each position within $C_i$.  Note that the subsequence DTW operations in this first step can be done in parallel.

The second step is to form a segment-level cost matrix $C_{seg}$.  In (regular) subsequence DTW, the optimal alignment path is found by identifying the lowest cost element in the last row of $D_i$, and then using $B_i$ to backtrack each step of the optimal path.  In WSDTW, however, our goal is to find an optimal global alignment path, not just a sequence of locally optimal alignment paths.  We construct $C_{seg} \in \mathbb{R}^{K \times M}$ by concatenating the last row of all matrices $D_i$, $i=1,2,\dots,K$.  In other words, $C_{seg}$ contains the optimal subsequence path scores ending at all possible locations.  The $C_{seg}$ matrix will play the same role as the pairwise cost matrix, but it describes costs at the segment level rather than at the frame level.
	
The third step is to find the optimal path through $C_{seg}$ using dynamic programming.  There are two types of allowable transitions.  The first transition is $(0,1)$ with weight $0$, which corresponds to skipping elements in the reference sequence with no penalty.  In Figure \ref{fig:overview}, this corresponds to moving directly to the right.  The second transition is $(1, \frac{N}{2K})$ with weight $1$, which corresponds to matching the shortest possible subsequence path across a chunk $C_i$ and transitioning to the next chunk $C_{i+1}$.  This dynamic programming formulation identifies the $K$ elements in $C_{seg} \in \mathbb{R}^{K \times M}$ that have the lowest cumulative cost while satisfying the following three constraints: (a) one element is taken from each row, (b) the elements are monotonically increasing, and (c) the selected path elements must be separated by at least $\frac{N}{2K}$ positions.  The optimal path can be found by performing dynamic programming on $C_{seg}$ to generate a cumulative cost matrix $D_{seg}$ and corresponding backtrace matrix $B_{seg}$, and then backtracking each step of the optimal path.  At the end of this third step, we have identified the ending locations in each $C_i$, $i=1,2,\dots,K$ of a series of subsequence paths that constitute a globally optimal path.

The fourth step is to backtrack through each of the frame-level cost matrices.  Using the optimal ending locations in each $C_i$, $i=1,2,\dots,K$, we use the information in $B_i$ to backtrack each step of the corresponding subsequence path.  The concatenation of the subsequence paths in $C_i$ forms our final estimate of the global alignment path.  Note that the ending location of each individual subsequence path is selected in a way that achieves a global optimal cost and ensures (weak) temporal consistency.

\subsection{Strictly-ordered Segmental DTW}
\label{subsec:ssdtw}

The second variant is called strictly-ordered Segmental DTW, which we abbreviate as SSDTW.  We will first provide a rationale for this variant, and then describe its differences from WSDTW.

One potential weakness of WSDTW is that it allows alignment paths that are not monotonically increasing.  To see this, note that WSDTW only imposes the constraint that consecutive elements in the optimal path through $C_{seg}$ be separated by a minimum distance $\frac{N}{2K}$.  While this ensures that all possible DTW paths are considered by WSDTW, it also introduces paths with backward discontinuities since the best subsequence paths through a chunk $C_i$ will almost certainly span a longer duration along the reference sequence than $\frac{N}{2K}$.  To address this issue, the SSDTW variant imposes additional constraints to ensure that alignment paths are monotonically increasing.  SSDTW has two main differences from WSDTW.

The first difference is that SSDTW constructs a segment-level transition matrix $T_{seg} \in \mathbb{Z}^{K \times M}$ in addition to $C_{seg}$.  This transition matrix keeps track of where subsequence paths start and end.  Each element $T_{seg}[i,j]$ specifies the starting location (along the reference sequence) of the best subsequence path in $C_i$ ending at position $j$.  Thus, constructing $T_{seg}$ means backtracking from every possible ending location and recording the starting location of each subsequence path.  The information in $T_{seg}$ will allow us to impose more specific constraints in the segment-level dynamic programming stage than WSDTW.

The second difference is in the segment-level dynamic programming stage.  There are two valid types of transitions to a position $(i,j)$ in $C_{seg}$.  The first transition is from $(i,j-1)$, which is the same skip as before.  The second transition is from $(i-1,T_{seg}[i,j]-1)$ with weight $1$.  Note that these transitions still allow for discontinuities in the forward direction (i.e.~a skip) at the boundaries of the chunks $C_i$, but they ensure that there are no backward discontinuities.  All other steps in SSDTW are the same as in WSDTW.

At a high-level, we can see that strictly-ordered Segmental DTW contains a substantial amount of additional computation in exchange for a guarantee of strict monotonicity in predicted alignment paths.

\section{Experimental Setup}
\label{sec:setup}

This section describes the setup of our empirical simulations.  The goal of these simulations is to characterize the behavior of the proposed alignment algorithms.

\begin{table}
\begin{center}
	\begin{tabular}{|l|c|c|c|c|c|} 
		\hline
		Piece & Files & mean & std & min & max \\
		\hline
		Opus 17, No 4 & 64 & 259.7 & 32.5 & 194.4 & 409.6 \\
		Opus 24, No 2 & 64 & 137.5 & 13.9 & 109.6 & 180.0 \\
		Opus 30, No 2 & 34 & 85.0 & 9.2 & 68.0 & 99.0 \\
		Opus 63, No 3 & 88 & 129.0 & 13.4 & 96.2 & 162.9 \\
		Opus 68, No 3 & 51 & 101.1 & 19.4 & 71.8 & 164.8 \\
		\hline
	\end{tabular}
\end{center}
\caption{Overview of the Chopin Mazurka data used in the alignment experiments.  All durations are in seconds.}
\label{tab:data}
\end{table}

We use the Chopin Mazurka dataset \cite{sapp2008hybrid} as a representative audio-audio alignment task.  Table \ref{tab:data} summarizes the data.  This dataset has been used as the basis for several studies on alignment and beat tracking \cite{schreiber2020modeling}\cite{grosche2010makes}\cite{sapp2008hybrid}, and provides reliable beat-level annotations for multiple performances of five different Chopin Mazurkas.  For each mazurka, we consider all pairs of performances and evaluate the accuracy of the predicted alignment.\footnote{We discarded a handful of queries (i.e. pairings of recordings) where the average time warping factor is greater than $2$.  In these cases, regular DTW (with the specified transitions) has no valid alignment paths, so we cannot measure its performance.}  For all experiments, we compute the pairwise cost matrix using a $L_2$-normalized constant-Q chromagram (23ms hop size) with cosine distance metric.  While specialized features have been proposed for various alignment tasks (e.g.~\cite{EwertMG09_HighResAudioSync_ICASSP}\cite{joder2013learning}), we stick with standard chroma features since our focus is on the alignment algorithm and not the feature representation.  We set aside one mazurka for debugging and development, and used the remaining four mazurkas as a test set.

We evaluate the alignment accuracy in the following manner.  For a given pair of recordings $A$ and $B$, we compare the predicted and ground truth timestamps in $B$ corresponding to the ground truth beat locations in $A$.  We consider an estimated beat location to be correct if the alignment error is less than a specified tolerance.  By sweeping across a range of tolerance values, we can characterize the tradeoff between error rate and tolerance.  The reported error rates on the test set are averaged across $7630$ queries (i.e.~pairs of recordings) and $1,930,922$ individual beat predictions.

\section{Results}
\label{sec:results}

\begin{figure}
	\includegraphics[width=\linewidth]{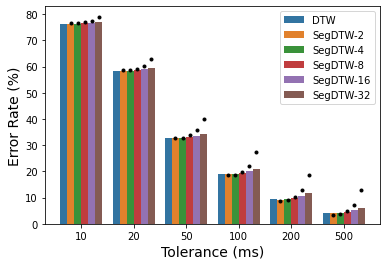}
	\caption{Comparison of DTW and both Segmental DTW variants on an audio alignment task.  All non-blue bars indicate the performance of WSDTW for different values of $K$ (amount of parallelization), and the black dots indicate the performance of SSDTW for the same values of $K$.}
	\label{fig:results}
\end{figure}

Figure \ref{fig:results} compares the alignment accuracy of DTW and the Segmental DTW variants.  Each group of bars corresponds to a different tolerance value.  Within each group of bars, the leftmost bar indicates the error rate of regular DTW, and the remaining bars indicate the error rate of weakly-ordered Segmental DTW for various values of $K$.  Recall that $K$ specifies the number of chunks to break the global cost matrix into, so it can be interpreted as the number of jobs to parallelize across.  The error rate of strictly-ordered Segmental DTW is indicated by black dots for the same set of $K$ values.  So, each histogram bar and the overlaid black dot indicate the performance of the two segmental DTW variants for the same value of $K$.

There are three things to notice about Figure \ref{fig:results}.  First, both segmental DTW variants match the performance of regular DTW for small values of $K$.  We observe this across the entire range of tolerance values.  Second, the performance of both segmental DTW variants gets worse as $K$ increases.  This is to be expected, since large values of $K$ mean that the subsequences get shorter and shorter and become less and less distinctive.  It is useful to point out that the effect of $K$ on alignment accuracy is data dependent.  On the Mazurka dataset, $K=32$ means that the subsequences for some pieces are less than three seconds long.  Third, the performance of WSDTW degrades much more gracefully than SSDTW as $K$ increases.  Contrary to our intuition, it seems that the allowance for occasional backward jumps is an asset, not a liability.  

Table \ref{tab:runtime} compares the runtime of all systems under controlled conditions.  We measure the wall clock time of DTW and the two Segmental DTW variants on (random) square cost matrices of increasing size.  All the times reported in Table \ref{tab:runtime} are for single-threaded optimized implementations in cython.  Even though the Segmental DTW algorithms can be parallelized, we profile single-threaded implementations for two reasons.  First, it allows for a direct comparison to DTW, making it clear how much total additional computation is required for the segment-level operations.  Second, it allows us to measure what fraction of the runtime is parallelizable in an environment-independent manner.  A parallelized implementation would incur additional runtime costs for setting up a parallelized computation, and the amount of this runtime cost will depend on the computing environment (e.g. distributed vs single server, hardware, job scheduling system, etc).  By measuring the runtime of individual stages of the Segmental DTW algorithms, we can estimate a bound on the amount of potential runtime savings through parallelization in a way that is largely environment-independent.  All experiments in Table \ref{tab:runtime} were done on a single Intel Xeon 2.1GHz CPU with 128GB of RAM, and each reported runtime is the average of 10 trials.

\begin{table}
	\begin{center}
		\begin{tabular}{|l|c|c|c|c|c|c|} 
			\hline
			System & 1k & 2k & 5k & 10k & 20k & 50k \\
			\hline
			DTW & .017 & .096 & .55 & 2.1 & 8.5 & 56.8 \\
			\hline
			WSDTW-2 & .020 & .088 & .57 & 2.2 & 8.6 & 54.2 \\
			WSDTW-4 & .019 & .092 & .62 & 2.3 & 8.6 & 54.1 \\
			WSDTW-8 & .020 & .091 & .50 & 2.2 & 8.7& 53.4 \\
			WSDTW-16 & .019 & .083 & .49 & 2.3 & 8.8 & 53.6 \\
			WSDTW-32 & .020 & .10 & .57 & 1.9 & 8.7 & 53.2 \\
			\hline
			SSDTW-2 & .026 & .12 & 1.0 & 4.2 & 17.8 & 112.0 \\
			SSDTW-4 & .025 & .12 & .86 & 4.3 & 18.9 & 121.0 \\
			SSDTW-8 & .026 & .13 & .67 & 3.1 & 17.6 & 125.3 \\
			SSDTW-16 & .026 & .12 & .67 & 3.2 & 12.7 & 125.5 \\
			SSDTW-32 & .035 & .14 & .70 & 2.6 & 12.8 & 86.1 \\
			\hline
		\end{tabular}
	\end{center}
	\caption{Comparison of average runtimes on cost matrices of different sizes (e.g. 5k indicates a $5000 \times 5000$ cost matrix).  Both WSDTW and SSDTW are evaluated with different values of $K$, but the algorithms are run on a single thread to compare the total amount of required computation.  All times are in seconds.}
	\label{tab:runtime}
\end{table}

There are two things to notice about Table \ref{tab:runtime}.  First, WSDTW has approximately the same average runtime as regular DTW, regardless of the value of $K$.  This indicates that the additional computation introduced by WSDTW does not significantly increase the total amount of computation.  Second, SSDTW requires roughly twice as much runtime as WSDTW.  This is the cost of calculating $T_{seg}$ to ensure monotonicity in the alignment.

\begin{figure}
	\includegraphics[width=\linewidth]{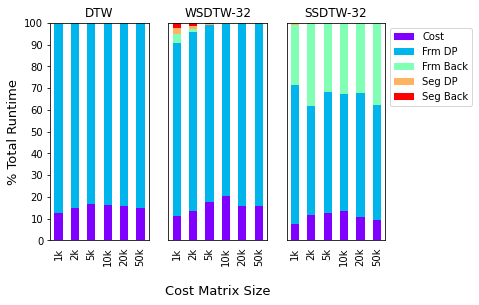}
	\caption{Breakdown of runtime by component.  Note that bar lengths indicate \textit{percentage} of total runtime.}
	\label{fig:runtimeBreakdown}
\end{figure}

Figure \ref{fig:runtimeBreakdown} shows the breakdown of runtime by component as the cost matrix size increases.  The runtime is broken down into five components: cost matrix computation (``Cost"), frame-level dynamic programming (``Frm DP"), frame-level backtracking (``Frm Back"), segment-level dynamic programming (``Seg DP"), and segment-level backtracking (``Seg Back").  We include the cost matrix computation in our measurements since a parallelized implementation of WSDTW or SSDTW would compute this in a distributed manner.  Note that the figure shows the \textit{percentage} of total runtime for each of the five components.

There is one key thing to notice in Figure \ref{fig:runtimeBreakdown}: nearly all of the computations in WSDTW and SSDTW are parallelizable.  The only components that are not parallelizable in Segmental DTW are the segment-level dynamic programming and segment-level backtracking.  For WSDTW, more than $99\%$ of the runtime is parallelizable for cost matrices of size $5000 \times 5000$ or greater.  In contrast, the only component in regular DTW that can be parallelized is the cost matrix computation, which accounts for $10-15\%$ of total runtime.

\section{Discussion}
\label{sec:discussion}

In this section we share three key insights that elucidate the relationship between DTW and the two Segmental DTW variants.

Insight \#1: WSDTW considers all DTW alignment paths.  More precisely, the set of all alignment paths considered by WSDTW is a superset of the alignment paths considered by regular DTW.  We can see this by noting that WSDTW allows for all valid DTW paths, and it also allows for sudden discontinuities at the boundaries of the subsequence chunks.

Insight \#2: SSDTW does \textit{not} consider all DTW alignment paths.  For a few problematic queries in the test set where there was severe time warping (average warping factor close to $2$), SSDTW did not have any valid paths through the cost matrix, though there were valid DTW paths.  This behavior arises because the segment-level transitions through each subsequence chunk cannot simply take (say) all (1,2) transitions; it is instead limited to the optimal subsequence paths through the cost matrix, which may or may not include paths containing all (1,2) transitions.  This phenomenon establishes that there are paths in DTW that are not considered by SSDTW.

Insight \#3: DTW and the Segmental DTW variants can produce completely unrelated paths on the same cost matrix.  We compared the predicted alignment paths of all three algorithms on randomly initialized cost matrices, where the best path is pure noise.  In these cases, DTW and the Segmental DTW variants generated completely unrelated paths that had no discernible resemblance to one another.  Often, SSDTW and WSDTW produced alignment paths that had many abrupt discontinuities, while DTW always produced a smooth alignment path (by constraint).

These three insights paint the following picture.  The set of SSDTW alignment paths is not guaranteed to contain the optimal DTW path, so it is not a suitable approximation for DTW.  Its use is not recommended.  The set of WSDTW alignment paths \textit{is} guaranteed to contain the optimal DTW path.  The size of the WSDTW set grows as $K$ increases, since it considers path discontinuities with increasing frequency.  How closely the optimal WSDTW path resembles the optimal DTW path depends on (at least) two key factors.  The first factor is the value of $K$, where larger values of $K$ will lead to a worse approximation.  The second factor is how distinctive subsequences in the data are, which is a characteristic of the data itself.  This can be thought of as a kind of signal-to-noise ratio (SNR) which describes how ``deep" the ravine is for the optimal path through the cost matrix.  For high SNRs, the optimal WSDTW path closely resembles the optimal DTW path.  For low SNRs, the optimal WSDTW path may not resemble the optimal DTW path at all.

\section{Conclusion}
\label{sec:conclusion}

We have examined two parallelizable alternatives to DTW for globally aligning two feature sequences: weakly-ordered Segmental DTW and strictly-ordered Segmental DTW.  Both alternatives break the global cost matrix into several sub-matrices, process the sub-matrices using subsequence DTW, and combine the results to find a globally optimal alignment path.  We find that WSDTW unilaterally outperforms SSDTW, and it matches the accuracy of DTW over a range of conditions.  Furthermore, nearly all of the operations in WSDTW are parallelizable.  Future work includes combining WSDTW with other cost-reduction methods like the Sakoe-Chiba band, characterizing the conditions under which WSDTW closely approximates DTW, and evaluating its performance on other alignment tasks.

\section{Acknowledgments}
This material is based upon work supported by the National Science Foundation under Grant No.~1948531.

\vfill\pagebreak

% References should be produced using the bibtex program from suitable
% BiBTeX files (here: strings, refs, manuals). The IEEEbib.bst bibliography
% style file from IEEE produces unsorted bibliography list.
% -------------------------------------------------------------------------
\bibliographystyle{IEEEbib}
\bibliography{SegmentalDTW}

\end{document}